\newcommand{\bm}[1]{\mbox{\boldmath $#1$}}
\documentclass[twocolumn,aps,showpacs,floatfix]{revtex4}

\usepackage{graphicx}
\usepackage{color}
\usepackage{epsf}

\usepackage{epsfig}
 
\begin{document}

\title{Inelastic Scattering from Core-electrons: a Multiple Scattering Approach}

\author {J.~A. Soininen,$^{1,2}$ A.~L. Ankudinov,$^2$ and J.~J. Rehr$^2$}
\affiliation{$^1$Division of X-ray Physics, Department of Physical Sciences,
University of Helsinki, FIN-00014 Finland}
\affiliation {$^2$Department of Physics, University of Washington,
Seattle, Washington 98195-1560}
\date{\today}

\begin{abstract}

The real-space multiple-scattering (RSMS) approach is applied to
model non-resonant inelastic scattering from deep core 
electron levels over
a broad energy spectrum.
This approach is applicable to aperiodic or periodic systems alike
and incorporates {\it ab initio}, self-consistent electronic structure
and final state effects.  The approach generalizes to finite momentum
transfer a method used extensively to model x-ray absorption spectra (XAS),
and includes both near edge spectra and extended fine structure.
The calculations can be used to analyze experimental 
results of inelastic scattering from core-electrons using either x-ray photons
(NRIXS) or electrons (EELS).
In the low momentum transfer region (the dipole limit), these inelastic
loss spectra are 
proportional to those from XAS. Thus their analysis can
provide similar information about the electronic and structural
properties of a system. Results for finite momentum transfer 
yield additional information concerning monopole, quadrupole,
and higher couplings.  Our results are compared both with experiment
and with other theoretical calculations.

\end{abstract}

\maketitle
\section{Introduction}

The primary aim of this work is to model inelastic scattering quantitatively
using a generalization of the real space multiple scattering (RSMS) approach.
This {\it ab initio} Green's function method has been extensively developed
for calculations of x-ray absorption spectra (XAS) and related
spectroscopies \cite{rehr00,alex98}. Thus the approach permits calculations of
inelastic scattering in arbitrary aperiodic materials over a broad energy
range, including the near edge and extended fine structure.
In inelastic scattering experiments, the scattering cross section is measured 
as a function of the energy $\omega$ and momentum ${\bf  q}$
transfered from the 
probe to the system (throughout this paper we use Hartree atomic units,
$e=\hbar=m=1$).
When the energy transfer is close to the binding energy of a 
core state, the inelastic scattering cross-section exhibits pronounced jumps,
from which one can study the core-excited states of the system.
We focus in this paper on the excitations from such core levels.

 In the low momentum transfer regime 
the scattering cross section is dominated by the dipole allowed excitations.
Qualitatively  the dipole approximation is valid when $qa \ll 1$, where
$a \sim 1/\zeta$ is the mean radius of the core state,  
$\zeta$ being the effective core charge. 
In the dipole limit the inelastic scattering
cross section is proportional to the x-ray absorption
coefficient~\cite{mizuno67,suzuki67,Ino71}, and
the direction $\hat{\bf q}$ of the momentum transfer vector plays the role
of the XAS polarization vector ${\hat{\bm \varepsilon}}$.
Thus the anisotropy of the x-ray edge can be studied by
varying $\hat{\bf q}$ in much the same way it is studied in XAS by varying
the direction of  ${\hat{ \bm \varepsilon}}$. However, as the magnitude of 
$q$ increases, contributions from other (dipole forbidden)
excitation channels become important.  The symmetries of a system
generally restrict the available excitation channels at a given excitation
energy.   The relative weights of the various excitation channels, as a
function of momentum transfer, also depend on the spatial extent of the
excited state. In general the classification of the spectral features
into given spatial symmetries of the final state electron
(i.e. $s$-type, $p$-type etc.) is not as straightforward as in XAS.
This is because contributions to the spectra from different excitations
often overlap at finite momentum transfer. Due to this
overlap, a detailed analysis of experimental results can be difficult
in the absence of quantitative theoretical calculations.
There are some exceptions though; 
core-excitons
can have well defined spatial symmetry and
energies well separated from other excitations.  
 
Non-resonant inelastic scattering spectra can be measured using
either electron energy-loss
spectroscopy (EELS)~\cite{ritsko1974} or non-resonant inelastic x-ray 
scattering (NRIXS)~\cite{doniach1971}. The cross sections in both 
cases are related to the dynamic structure factor 
$S({\bf q},\omega)$ of a system. For the case of x-rays, the non-relativistic 
Born approximation for the NRIXS double differential cross-section 
is~\cite{Schulke1991,Hamalainen2001} 
\begin{equation} \label{NRIXS}
\frac{d^2 \sigma}{d \Omega d \omega} = \left ( \frac{d \sigma}{d \Omega}
\right) _{Th} S({\bf q},
\omega).
\end{equation}
Here $(d\sigma/d\Omega)_{Th}$ is the Thomson scattering cross-section.
The Thomson cross section can be expressed in terms of the incoming (scattered) photon 
polarization vector and energy  
$\{{\hat{\bm\varepsilon}_1},\omega_1\}$ 
($\{{\hat{\bm\varepsilon}_2},\omega_2\}$) as 
\[
\left(\frac{d\sigma}{d\Omega}\right)_{Th}=r_0^2
({{\hat{\bm \varepsilon}}_1 \cdot {\hat{\bm \varepsilon}}_2})^2 \frac{\omega_2}{\omega_1},
\]
where $r_0$ is the classical electron radius. 
In Eq.~(\ref{NRIXS}) we have implicitly assumed that the energy of the incident 
x-ray is not close to any core binding energy $E_i$ (typically $E_i\ll \omega_1$). 
This is in contrast with resonant inelastic x-ray scattering experiments where the 
incident x-ray energy is tuned close to a core binding energy and resonant processes dominate 
the double differential cross section. The different forms of inelastic x-ray scattering
and their relationships are discussed in more detail in Refs.~\cite{Schulke1991,Hamalainen2001}.  
In EELS the corresponding scattering cross section in the Born approximation 
is~\cite{Jones73} 
\begin{equation}
\frac{d^2 \sigma}{d \Omega d \omega} = \left ( \frac{d \sigma}{d \Omega} 
\right) _{e-e} S({\bf q},
\omega),
\end{equation}
where $({d \sigma}/{d \Omega})_{e-e}= q^{-4} (p'/p) $ is the electron-electron 
scattering cross-section with $p$ ($p'$) being the momentum of the
incoming (scattered) electron. In other words, with these approximations 
the cross-sections differ only by a coupling constant which is
related to the probe (i.e., a photon or electron) that interacts with 
the system. The system dependent part, namely the dynamic structure factor
$S({\bf q}, \omega)$,
is the same for both NRIXS and EELS, and depends only on the structure and
excited state properties of the system under investigation.

Non-resonant inelastic x-ray scattering from core-excited states
is often called x-ray Raman scattering (XRS). 
Traditionally XRS experiments have been limited to $K$-edges of low-$Z$ materials. 
However, for the $K$-edges, experimentally 
accessible systems include such important cases as those with carbon
and oxygen, as has already been demonstrated (see Ref. \cite{bergmann2002} and 
references therein). 
Recently XRS experiments were done for 
the $L$-edges and $N$-edges of materials with higher $Z$ 
including N-edge of Ba in Ba$_8$Si$_{46}$~\cite{sternemann05}. 
The smallness of the x-ray scattering cross-section implies
that bulk information is generally obtained in these probes, i.e., that 
multiple scattering of the probe particles and 
surface effects are usually not serious problems. 
XRS has been used to study both the
momentum transfer magnitude~\cite{nagasawa1989,krisch1997,soininen2001,hamalainen2002,feng04}
and direction dependence~\cite{nagasawa1989} at $K$-edges.
Recently the $K$-edge of Be metal was studied using both degrees of freedom
(magnitude and direction)~\cite{sternemann03}.  
The typical energy resolution of current XRS experiments is around
1.0~eV FWHM, although some experiments with a energy resolution of 0.4~eV
have been carried out~\cite{krisch1997}.  
The finite core-hole lifetime broadening also limits the useful energy
resolution to a few tenths of an eV. 

As noted above, one can also measure the fine structure of the
spectra above deep core edges using electrons as a probe.
Unlike non-resonant x-ray scattering, EELS is not limited to low-$Z$
materials, and has a typical energy resolution on the order of 0.1 eV.
Electrons interact strongly with solids, so that one is often able
to obtain results with good statistical accuracy. However, their strong
interaction also implies that multiple scattering of the probe electrons
can be a problem which has to be removed or otherwise accounted for in
the analysis of experimental results. This is especially true of experiments
with large scattering angles or low incident electron energies,
in which case one should explicitly include multiple scattering in
the analysis, as explained in Ref.~\cite{Tomel89,Fujikawa02}.
The dependence of x-ray edges on the magnitude of the momentum transfer
has been investigated in Ref.~\cite{ritsko1974,Fie77,Hitchcock2000}, and more recently in
Ref.~\cite{Jiang04}. The dependence on the momentum transfer direction 
at x-ray edges has also been used in the analysis of anisotropy of 
x-ray edges (see Ref.~\cite{Klie03,Jiang02} and references therein).
Additionally 
the very high spatial resolution available in EELS makes it possible to 
study the anisotropy of x-ray edges in fine-grained solids~\cite{Jiang02}.

Theoretically the problem of calculating inelastic scattering 
spectra from core-excited states is analogous to the 
calculation of core level x-ray absorption, the main difference being in the
nature of the transition matrix elements. Thus there are number of
effects that must be taken into account for accurate computational results.
Within the independent electron approximation,
the main problem is the behavior of the core-excited final states. This can
be approached either by calculating the core-excited states one by one
(e.g., in band-structure or molecular-orbital based approaches)
or equivalently, by calculating the one-electron Green's function
for the excited states, as is done in this work.
For accurate results one needs a highly quantitative method for calculating
single-particle properties such as electron densities and Coulomb
potentials. Also a number of many-body properties of the
excited states of the system have to be considered. These include the 
interaction of the final state electron with the core-hole created in the
excitation process, as well as an approximation for the energy-dependent
self-energies $\Sigma(E)$ of both the electron and the core-hole, e.g.
to account for inelastic losses.
A good approximation in many cases is to calculate the final
state in the presence of the core hole and the initial state
with ground state potentials \cite{rehr00}.  This approximation is
referred to as the {\it final state rule}.  
In principle one should include the whole energy dependence of the 
self-energy, as discussed for example in Ref.~\cite{campbell02}. 
However, often a quasiparticle approximation is used both for the electron 
and the core-hole.
In the quasiparticle approximation, the imaginary part of the self-energy
causes the electron (and the hole) in the excited state to have a finite
lifetime.  Additionally
the real-part of the self-energy shifts the spectral features from 
the positions predicted by ground-state (or mean-field) approximations,
whereas traditional ground state density functional treatments can lead
to significant errors in peak positions and intensities.  For a full
treatment, especially at high energies, one also needs
to consider coupling to vibrational degrees of freedom, which give the
spectra a temperature dependent damping comparable to that of x-ray
Debye-Waller factors.  A more detailed review of the problem of  
computing core-excited states in XAS can be found in Ref.~\cite{rehr00}.
As noted above the main difference between inelastic scattering and
x-ray absorption lies in
the transition matrix elements. This is due to the fact that the
operator mediating the transitions in 
inelastic scattering is $\exp( i{\bf q}\cdot {\bf r})$, as compared to the
dipole operator ${\hat{ \bm \varepsilon}}\cdot{\bf r}$ 
(and, in some cases, also the quadrupole operator)
in x-ray absorption. We will discuss this difference in more detail below.

In this article we have applied the real-space multiple-scattering (RSMS)
approach to calculate the core-excited states
in EELS and NRIXS over a wide energy range.
The RSMS approach adopted here, is essentially an effective, independent
particle approach for excited states which takes into account the most
important final state effects. The approach is an extension of the
RSMS approach previously used for x-ray absorption spectra~\cite{rehr00}.
The momentum transfer dependence of both the near edge
spectra and the fine structure in inelastic loss spectra
are analyzed in detail and compared with
experimental results.

Recently a two-particle approach to core-excited states has been
developed~\cite{shirley1998,soininen2001a} based on the
Bethe-Salpeter equation (BSE). The BSE goes beyond the independent particle
approximation with an explicit treatment of the screened electron core-hole
interaction. However, when the core-hole interaction is strong, the
results from the final state rule can be comparable \cite{rehr04}.
The BSE approach has been applied to different situations using
the momentum transfer dependent matrix elements in the analysis of 
XRS~\cite{soininen2001a,soininen2001,hamalainen2002,sternemann03}.
For completeness our calculations are also compared with
calculations based on the BSE.
However, the BSE approach is currently applicable only relatively close to
a given edge, typically between a few tens and 100 eV depending on the
material, and becomes computationally impractical for
treating the extended fine structure in inelastic scattering spectra.

\section {Real Space Multiple-scattering formalism}

\subsection{Green's function formulation}

We now briefly describe the extension of the real-space
multiple-scattering 
approach for calculations of
non-resonant inelastic scattering.  The RSMS approach can be regarded as the
real space analog of the KKR band structure method \cite{schaich73}, but
unlike KKR, the method makes no assumption of symmetry and is applicable to
periodic and
aperiodic systems alike.  Moreover, the implementation used here is an
{\it all-electron} approach which can be applied to arbitrary systems
throughout the periodic table.  This method has been extensively developed
for {\it ab initio} calculations of x-ray spectra including
both EXAFS (extended x-ray absorption fine structure) and XANES
(x-ray absorption near edge structure) \cite{rehr00,alex98}.  Its generalization
here to finite momentum transfer ${\bf q}$ is relatively
straightforward, though non-trivial in several respects, as discussed
below. 
The contributions to $S({\bf q},\omega)$ from excitations from a tightly bound 
core state $| i \rangle$ can be approximated using Fermi's Golden rule, i.e.,
\begin{equation}
S_i({\bf q},\omega) = \sum_f |\langle f | e^{i{\bf q}\cdot {{\bf r}}} | i \rangle|^2
                    \delta(\hbar \omega+ E_i-E_f),
\end{equation}
where $E_i$ ($E_f$) is the initial (final) state quasiparticle energy
of the electron. For the remainder of the paper we will drop the 
index $i$ and use simply $S({\bf q},\omega)$ when referring to
the contribution from a core state $| i \rangle$. 
Within the one-electron approximation and the final state rule,
the final (photo-electron) states $|f\rangle$
are quasi-particle states which are eigenfunctions of the final state
Hamiltonian in the presence of an adiabatically screened core hole (denoted
with a prime), i.e., $H'= p^2/2m + V'_{\rm coul} + \Sigma(E)$.
Here $\Sigma(E)$ is the self-energy (or dynamically screened
exchange-correlation potential) which accounts for inelastic losses, which
are essential for a proper treatment of inelastic electron scattering.  In our
calculations we use the energy
dependent local density approximation for $\Sigma(E)$ of Hedin and
Lundqvist~\cite{Hedin71},
based on the plasmon-pole approximation for the dielectric function. 
This approximation is usually adequate at moderate to high photoelectron
energies \cite{rehr00}.  The states $|i\rangle$ are deep-core levels
of the initial state Hamiltonian without a core-hole.
For  small momentum transfers in the case of core-excited states, 
the excitation operator $\exp(i{\bf q}\cdot{\bf r})$ can be expanded as
\begin{equation}
\exp(i{\bf q}\cdot{{\bf r}}) \approx 1+i{\bf q}\cdot {\bf r} + O(q^2).
\end{equation}
Thus the dipole selection rule is approximately valid at small $q$, since
the first term should not contribute to transitions due to orthogonality of
the initial and final states, and the next term ($\propto q^2$) is also small.
In fact since $\omega$ and ${\bf q}$ can, in principle,
be chosen separately, the dipole approximation can sometimes be satisfied
better in inelastic scattering than 
in absorption, where the photon momentum and energy are always linked.
At higher $q$, dipole forbidden transitions become important,
and it is more convenient to expand the exponential in terms of spherical 
harmonics
\begin{equation}
\exp(i{\bf q}\cdot{\bf r}) = 4\pi \sum_{lm} i^l j_l(qr) Y^*_{lm}({\bf \hat q})
Y_{lm}({\bf \hat r}).  
\end{equation}
Thus for any finite momentum transfer all excitation
channels are present.

To avoid explicit calculations of final states it is advantageous
to  re-express the Golden rule in terms of the one-particle
Green's function or propagator $G=[E-H'+i\Gamma]^{-1}$
in real space, where $\Gamma$ is the core-hole lifetime.
Thus using the spectral representation, $ -(1/\pi) {\rm Im}\, G(E) =
              \Sigma_f |f\rangle\langle f| \delta(E-E_f) \equiv \rho(E)$
the dynamic structure factor becomes
\begin{equation}
 S({\bf q},\omega) = 
   \langle i | e^{-i{\bf q}\cdot {\bf r'}} P \rho({\bf r'},{\bf r},E)
   P e^{i{\bf q}\cdot {{\bf r}}} | i\rangle,
   \label{eq:golden}
\end{equation}
where the photoelectron energy is $E=\omega+E_i$. The operator $P$
projects the Green's function on to the unoccupied states of the initial
state (without the core hole) Hamiltonian~\cite{campbell02}. The 
operator $P$ is needed here since, in general, the eigenstates of the final 
state Hamiltonian are not strictly orthogonal to the initial state
of the scattering.
In practice this non-orthogonality implies that at low momentum transfer 
\begin{equation}
\langle i |e^{-i{\bf q}\cdot {\bf r}} | f \rangle \approx \langle i | f
\rangle -i \langle i |{\bf q}\cdot{\bf r} | f \rangle+\cdots .
\end{equation}
Thus the dipole forbidden excitations, which incorrectly vary as
$\propto \langle i | f \rangle$,
can begin to dominate over the dipole allowed transitions $\propto q$
as $q\rightarrow 0$. The application of the full 
projection operator $P$ would require taking an inner-product and appropriate 
matrix elements of all the occupied states $| j \rangle$ of the
initial state Hamiltonian, i.e.,
\begin{equation}
Pe^{i{\bf q}\cdot {\bf r}}  | i \rangle = e^{i{\bf q}\cdot {\bf r}}| i \rangle - \sum_j^{occ} | j \rangle \langle j |  e^{i{\bf q}\cdot {\bf r}} | i\rangle.
\end{equation}
Instead we simply approximate the effect of $P$ by modifying the excitation
operator as
\begin{equation}
\tilde{P}e^{i{\bf q}\cdot {\bf r}}  | i \rangle \approx
[e^{i{\bf q}\cdot {\bf r}}- \langle i | e^{i{\bf q}\cdot {\bf r}} | i \rangle ]
 | i \rangle.
\end{equation}
Clearly at low momentum transfers this gives matrix elements 
\begin{equation}
\langle i |e^{-i{\bf q}\cdot {\bf r}} \tilde{P} | f \rangle
\approx -i\, \langle i |{\bf q}\cdot{\bf r} | f \rangle+ \Theta (q^2),
\end{equation}
while at high momentum  transfers
the correction $\langle i | e^{i{\bf q}\cdot {\bf r}} | i \rangle  \langle f
| i \rangle $ approaches  zero. Although in what follows we do not
explicitly include the operator $\tilde{P}$, it is included 
in the actual calculations.  
The terms coming from the difference of $P$ and $\tilde{P}$ are neglected. 
Although this is a somewhat uncontrolled approximation, comparison 
with experiments and previous XRS calculations at different momentum 
transfers show that these terms should be small, as illustrated in
Fig.~\ref{EvsF} below.
Also work in XAS suggests that the difference not significant compared with
other theoretical uncertainties \cite{campbell02}.

In multiple scattering (MS) theory,
the scattering perturbation is the total electron potential,
which 
is separated into contributions from ``scattering potentials" $v_R$
which are localized on each atomic site $\bf R$, i.e.,
\begin{equation}
V'_{\rm coul}+\Sigma(E) = \sum_R v_R({\bf r}-\bf R).
\end{equation}
In the RSMS method $v_R({\bf r})$ is usually taken to be spherically
symmetric. This is a good approximation for electron scattering
calculations at moderate electron kinetic energies, i.e.,   
a few eV above the threshold.
These potentials are calculated self-consistently by
iterating the total electron density, potential and Fermi-energy,
typically requiring about 10-20 iterations. Once the potentials are known,
scattering phase shifts $\delta_l$ are calculated
and dimensionless $t$-matrices evaluated using the 
relation $t_l=\exp(i\delta_l)\sin(\delta_l)$.
With spherically symmetric potentials, the propagator $G(E)$ (and hence the 
density matrix $\rho(E)$) can be represented in an
angular-momentum $L=(l,m)$ and site ${\bf R}$ basis $|L,{\bf R}\rangle$.
Thus at the site of the core-excited atom (${\bf R}={\bf 0}$)
\begin{equation}
\rho({{\bf r'},\bf r},E)=\sum_{L,L'}
R_L({\bf r},E)\rho_{L,L'}(E) R_{L'}({\bf r'}, E),
\end{equation}
where $R_L({\bf r},E)$ are scattering states at energy $E$,
which are regular at the origin. The expansion for ${\bf r}$ and 
${\bf r'}$ about different sites ${\bf R}$ and ${\bf R'}$
is similar, with $G_{L,L'}(E)$ replaced by 
$G_{L{\bf R},L'{\bf R'}}(E)$.
Consequently the calculation of the dynamic structure factor is
reduced to a calculation of (embedded) atomic transition matrix elements
$M_L({\bf q}, E)=\langle R_L(E)| e^{i\bf q\cdot {\bf r}} |i\rangle$
and a multiple-scattering matrix $\rho_{L,L'}(E)=
\rho_{L{\bf 0},L'{\bf 0}}(E)$, i.e.,~\cite{qalongz} 
\begin{equation}
   S({\bf q}, \omega) = \sum_{LL'} M_{L}(-{\bf q}, E)  \rho_{L, L'} (E)
    M_{L'}({\bf q}, E).
\label{offdiagrep}
\end{equation}
Here $\rho_{L, L'} (E) = (-1/\pi){\rm Im}\, G_{L,L'}(E)$ denote matrix
elements of the final state density matrix, including the effect of
the core-hole potential.
It can be shown that $S({\bf q}, \omega)$ also satisfies a generalized
oscillator strength sum-rule.
The representation in
Eq.~(\ref{offdiagrep}) shows that the essential physics of the
problem separates into two parts: i) a ${\bf q}$ dependent transition matrix
which governs the production of photoelectrons into various final
states, and ii) a propagator matrix $G_{L,L'}$ which describes the scattering
of the photoelectron within the system at a given excitation energy. 
The transition matrix elements $M_{L}({\bf q}, E)$ are calculated using the
expansion in Eq.~(5) 
of  $\exp(i{\bf q}\cdot{\bf r})$ in terms of spherical harmonics.
Depending on the excitation energy and momentum transfer, different
terms $L$ in this expansion are important. For low momentum 
transfer and excitation energy the small $\Delta l $ transitions 
are most important, starting from the dipole $\Delta l =\pm 1$, 
monopole $\Delta l =0$, and quadrupole $\Delta l =\{\pm 2, 0\}$
transitions. Thus provided one can neglect the coupling to phonons (which is 
a good approximation for core excitations discussed in this work),
a typical spectrum at low $q$ can be analyzed using only 
these three excitation channels.
Monopole transitions, which are
present in $S({\bf q}, \omega)$ through the  term
$\propto 4\pi j_0(qr) Y^*_{00}({\bf \hat q}) Y_{00}({\bf \hat r})$,
have no counter-part in absorption 
($s$-to-$s$ transitions being forbidden within the dipole approximation).
When the momentum 
transfer is increased in inelastic scattering, other excitation channels 
become more important, especially at high energy values. 

The above RSMS formulation can be advantageous even for crystals,
since periodicity is broken by the core-hole interaction and
spectral broadening from the core-hole lifetime and the
self energy. Typically $\Gamma_{ch} + |{\rm Im}\, \Sigma|$ is
several eV at photoelectron energies above about 30 eV of an edge.
This broadening limits the
range probed by the photoelectron to clusters of order a few hundred 
atoms, and gives rise to a {\it short range order} theory for energies
above about 30 eV. Thus long range effects such as the sharp van Hove
singularities of band structure calculations are naturally smeared out.
 Moreover, in this extended energy regime
scattering is relatively weak and perturbation theory converges well.
Conversely, in the near edge regime (energies less than about 30 eV), 
the range is dominated mostly by the core lifetime $1/\Gamma_{ch}$
which is very long for low $Z$ materials or for very shallow edges. For those
cases an intermediate- or long-range order theory may be needed,
involving multiple-scattering to all orders 
or very large cluster of atoms.

 For polycrystalline materials or systems with cubic symmetry, only
the diagonal terms in $L$ and $L'$ survive in Eq.~(9), corresponding
to couplings to various partial local, projected densities of states
(LDOS) $\rho_l(E)$,  i.e.,
\begin{equation} \label{diagn}
S({\bf q},\omega) \approx \sum_{l} (2l+1) |M_{l}({\bf q}, E)|^2  \rho_{l} (E).
\label{diagrep}
\end{equation}
Thus the dynamic structure factor is directly related to the LDOS.
The coupling terms $M_{l}({q}, E)$ are essentially atomic
quantities which can be calculated theoretically to good accuracy.
Thus it may be possible to extract the LDOS $\rho_{l} (E)$ or
the density matrix components $\rho_{L,L'}(E)$ from  a sequence of
experimental measurements of $S({\bf q},\omega)$ at various ${\bf q}$
by a suitable inversion procedure of Eq.~(\ref{offdiagrep}) or
(\ref{diagrep}), regarded as a set of linear equations.

The propagator $G=G^c+G^{sc}$ naturally separates into intra-atomic
contributions from the central atom $G^c$ and from 
MS contributions from the environment $G^{sc}$. 
Thus as in XAS,
$S({\bf q}, \omega)$ can be factored as 
\begin{equation}
S({\bf q},\omega) =S_0(q,\omega)[1+\chi_{\bf q}(k)],
\end{equation}
where $S_0(q,\omega)$ represents a smoothly varying atomic background, and
$\chi_{\bf q}(k)$ is the ``fine structure" due to multiple-scattering from the
environment. Here we have also used a standard XAS notation, where
$k=\sqrt{2(\omega+E_i)}$ is the photoelectron wave-number.
The momentum transfer dependence of 
these two contributions to the dynamic structure factor can 
be analyzed separately as shown in detail below.

\subsection{Central Atom Contribution}

Much of the qualitative behavior of the 
spectra can be understood in terms of the central atom contribution (in
the absence of other scatterers), which 
has been analyzed for isolated atoms by, e.g., Leapman et 
al.~\cite{Leapman79}. This contribution is independent of the direction
of the momentum transfer, and hence depends only on the magnitude $q$.
For a condensed system, the states $R_L({\bf r},E)=i^l R_l(r)Y_L(\hat r)$,
where $R_l(r)$ are radial wave functions and $Y_L(\hat r)$ spherical harmonics,
are scattering states defined for the potential of an ``embedded" atom in
the system. However, the deep core states differ little from those of isolated
atoms and spherical symmetry is still a good approximation.
This yields the atomic background contribution
\begin{equation}
S_0(q,\omega) = \sum_{l}(2l+1) |M_{l}(q, E)|^2  \rho^0_{l} (E),
\end{equation}
where $\rho^0_l$ denote the diagonal matrix elements of the density matrix.

\begin{figure}
\includegraphics[width=1.00\linewidth]{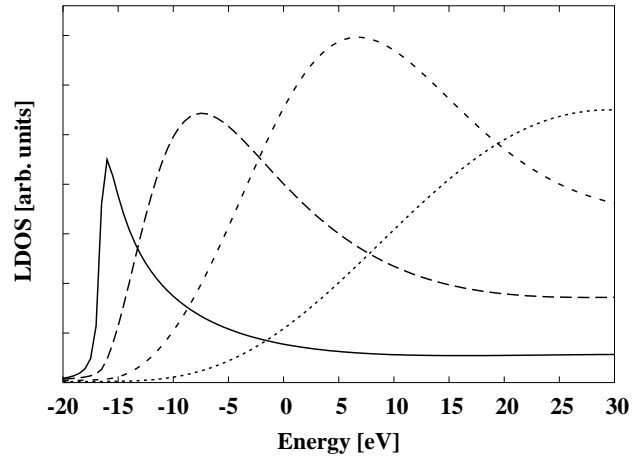}
\caption{LDOS $\rho^0_l(E)$ for the central site of fcc Al, in the absence
of other scatterers: $l=0$ (solid); $l=1$ (long dashes); $l=2$ (dashes); and
$l=3$ (short dashes). Note how the successive terms $\rho^0_l(E)$ with 
increasing angular momentum $l$ 
peak at successively higher energies and overlap each other.} 
\label{AlLDOS}
\end{figure}

The matrix elements can be calculated by rotating the system so that 
the momentum transfer is along $z$-axis. Additionally summing over the $m$
quantum numbers of both the initial state and the final state partial wave and 
using properties of 3-$j$ symbols one finds 
\begin{eqnarray}
|M_l(q, E)|^2 &=&(2l_i+1) \sum_{l'} (2l'+1)
\left | {\left( \begin{array}{ccc} l_i & l' & l \\ 0 & 0 & 0 \end{array}
\right) } \right. \times \nonumber \\ 
&\times& \left. {\int r^2 dr R_l(r,E) j_{l'}(qr) R_i(r) } \right|^2.
\end{eqnarray}
Since $j_{l'}(qr) \approx (qr)^{l'}$ near the origin,
the dominant term for small $q$ is the dipole approximation $l=1$.
The first term in a power series expansion of $j_0(qr)$ 
cancels by orthogonality for $q\rightarrow 0$, and hence the first
contributing terms of $j_{l'}(qr)$ from both $l'=0$ and $l'=2$
behave as $(qr)^2$. For large momentum transfers
such that $qa>1$ and higher excitation energies, successively larger angular
momentum ($l'>2$) excitation channels become more important.  As a consequence,
the behavior of $S_0(q,\omega)$ at increasing $q$ reflects that of the LDOS
of increasing $l$.
This is illustrated in Fig.~\ref{AlLDOS} for the background contribution
$S_0({\bf q},\omega)$.
For EELS however, experimental measurements of the double differential
cross-section $d^2 \sigma/d\Omega d\omega$ still tend to favor the
dipole-approximation due to the $q^{-4}$ dependence of the electron
scattering cross-section $(d\sigma/d\Omega)_{e-e}$.
Thus most of the experimental signal in EELS is at small $q$. 

\subsection{Fine Structure}

>From Eq.~(9) and (11), the normalized fine structure in the dynamic structure
factor is given by
\begin{equation}
   \chi_{\bf q}(k) =\frac{1}{S_0({\bf q},\omega)}
   \sum_{LL'} M_{L}({-\bf q}, E)  \rho^{sc}_{L, L'} (E) M_{L'}({\bf q}, E).
\label{chidef}
\end{equation}
where $\rho^{sc}_{L, L'} (E) = (-1/\pi){\rm Im}\, G^{sc}_{L,L'}(E)$.
Clearly the directional dependence of the spectra on $\hat q$ comes only
from the scattering contribution, and naturally $\chi_{\bf q}(k)$
will change with the magnitude of the momentum transfer due to the
${\bf q}$ dependence of the coupling terms.
This momentum transfer dependence of the near edge structure has been
used to study the symmetries of the core excited states in
solids~\cite{nagasawa1989,krisch1997,soininen2001,hamalainen2002,feng04,sternemann03,ritsko1974,Fie77,Jiang04,Klie03,Jiang02}. It may also be possible to
use this momentum transfer dependence to obtain information about
intermediate- and long-range structure, which are contained in the EXAFS-type
oscillations $\chi_{\bf q}(k)$.

One way of calculating the scattering contribution to the propagator
$G^{sc}$ for a big cluster of atoms is to use matrix
inversion involving the free particle propagator 
$G^0$ and the scattering matrix $T$, i.e.,
\begin{equation}
G^{sc}_{L,L'}= e^{i \delta_l} [({\bf 1} - G^0 T)^{-1} G^0]_{L,L'}
               e^{i \delta_{l'}}.
\end{equation}
This representation is referred to as full multiple scattering  (FMS),
since it formally includes MS to all orders. FMS is often used to
treat near-edge spectra where scattering is strong and the dimensions
of the matrix $G_{L,L'}$ are relatively small.  At higher energies or whenever
the MS series converges well, the matrix $G_{L,L'}$ can alternatively
be calculated in terms of a ``path expansion," i.e., as a sum over all
MS paths that a photoelectron can take away from the absorbing atom and
back \cite{lee75}.  Formally the path expansion is given by the sum
\begin{equation}
  G^{sc}_{L,L'} =  e^{ i \delta_{l}} [ G^0 T G^0 +
  G^0 T G^0 T G^0 + \cdots ]_{L,L'} e^{ i \delta_{l'}},
\end{equation}
where the successive matrix products terms represent single, double,
and higher order scattering processes.
Remarkably the path expansion has been found to be generally adequate
for energies above about 30 eV of threshold, where of order 10$^2$
of the largest amplitude paths suffice to yield an accuracy of a few percent
for most materials \cite{rehr90}.

Due to the large dimension $D\approx N(l_{max}+1)^2$
of the matrix $G_{L{\bf R},L'{\bf R}'}$,
where $N$ is the size of the cluster, exact
calculations with the path expansion can only be carried out for a few,
low-order MS paths. To overcome this bottleneck, an exact separable
representation
of the propagator was introduced by Rehr and Albers (RA) \cite{rehr90},
\begin{equation}
G^0_{L'{\bf R'},L''{\bf R''}}={e^{ikR}\over kR}
\sum_{\lambda}Y_{L',\lambda}(k{\bf R'})\tilde Y_{\lambda,L''}(k{\bf R''}),
\label{g0RA}
\end{equation}
where ${R}=|{\bf R'}-{\bf R''}|$ is the bond distance.
The generalized spherical
harmonic expansion coefficients $Y_{L',\lambda}$ and
$\tilde Y_{\lambda,L''}$ converge rapidly in
powers of $1/kR$, which is always $\ll$ unity for unoccupied states 
$k>k_F$, and hence this representation can be severely
truncated. For FMS calculations the RA representation of the propagators
in Eq.~(\ref{g0RA}) is stable and converges rapidly. For the path
expansion, the RA approach is exact for
single scattering, and typically only six terms in $\lambda$ suffice to give
accuracies to of order a percent or better over the typical range of
wave numbers encountered in XAS experiment,
$k_F\leq k \leq  20$ \AA$^{-1}$, i.e., $E_F \leq E \leq 1500$ eV.

With the separable representation, one can sum over all intermediate
angular momentum variables $(l,m)$ at each site and define local
{\it scattering matrices}.
Then the contribution $ G^0TG^0T \cdots G^0$
to the total propagator from a given $N$-atom path
$\Gamma\equiv [{\bf R}_1, {\bf R}_2, \dots {\bf R}_N={\bf R}_0]$
can be factored as a matrix product over  small (typically
$6\times 6$) 
matrices~\cite{rehr90}
\[
F_{\lambda,\lambda'}({\bm \rho}, {\bm \rho}')= \sum_L  \tilde{Y}_{\lambda,L}({\bm \rho }) t_l Y_{\lambda',L}({\bm \rho }'),
\]
which is the analog of the
scattering amplitude $f(\theta)$ in plane-wave scattering theory, and a
termination matrix
$m_{L,L'}^{\lambda_1,\lambda_N}({\bm \rho}_1, {\bm \rho}_N)
  =Y_{L,\lambda_1}({\bm \rho}_1)
                           \tilde Y_{L_N,\lambda_N}({\bm \rho}_N)$, where
${\bm \rho}=k{\bf R}$ \cite{rehr90}.
For large $L$ or $L'$, it may be necessary to increase the dimension of
$\lambda$ in the termination matrices $m_{L,L'}^{\lambda_1,\lambda_N}$
beyond 6 for adequate convergence.  Specifically we obtain for each path
(cf. Ref.~\cite{rehr90}),
\begin{eqnarray}
G^{\Gamma}_{L,L'} &=& \frac{e^{i(\rho_1+\rho_2+\cdots
\rho_N+\delta_l+\delta_{l'})}}
                    {\rho_1 \rho_2 \cdots \rho_N} 
\sum_{\{\lambda\}}  m_{L,L'}^{\lambda_1,\lambda_N}({\bm \rho}_1, {\bm \rho}_N) \times
     \nonumber  \\ 
 &\times& \left[ { F_{\lambda_N,\lambda_{N-1}}({\bm \rho}_N, {\bm \rho}_{N-1}}) 
  \cdots 
          F_{\lambda_2,\lambda_1}({\bm \rho}_2, {\bm \rho}_{1})  \right]. 
\end{eqnarray}
In our code, the dependence on the bond vectors ${\bm \rho}$ and ${\bm \rho'}$
is simplified using rotation matrices and Euler angles.
This expression is the similar to one used for analysis of 
XAS using path expansion~\cite{rehr00}. 
In contrast to the case for XAS which is dominated
by the dipole-approximation, however, all angular momentum channels
now contribute with ${\bf q}$ dependent couplings.
Thus the fine structure from the path expansion can be written as
\begin{eqnarray}
\chi_{\bf q}(k)&=& -{\rm Im} \sum_\Gamma
\frac{e^{i(\rho_1+\rho_2+\cdots \rho_N)}}
                    {\rho_1 \rho_2 \cdots \rho_N} 
\sum_{\{\lambda\}} H^{\lambda_1,\lambda_N}({\bm \rho}_1,{\bm \rho}_N,{\bf q})
    \nonumber \\ 
      &\times& \left[ { F_{\lambda_N,\lambda_{N-1}}({\bm \rho}_N, {\bm \rho}_{N-1}) 
  \cdots 
          F_{\lambda_2,\lambda_1}({\bm \rho}_2, {\bm \rho}_{1}) } \right]. 
\label{newform}
\end{eqnarray}
where
\begin{eqnarray}
&& H^{\lambda_1,\lambda_N}({\bm \rho}_,{\bm \rho}_N,{\bf q}) =
\frac{1}{S_0(q,\omega)}\sum_{L,L'} e^{i(\delta_l+\delta_{l'})} \times
\nonumber \\
&&\times \qquad M_{L}(-{\bf q})
         m_{L,L'}^{\lambda_1,\lambda_N}
({\bm \rho}_1, {\bm \rho}_N) M_{L'}({\bf q}) . 
\end{eqnarray}
Note that only the coupling terms
$H^{\lambda_1,\lambda_N}({\bm \rho}_1,{\bm \rho}_N,{\bf q})$ depend on
${\bf q}$, while the scattering contribution is a product of low order
scattering matrices $F_{\lambda,\lambda'}$ which is dependent only on
the material.
When the dominant contribution comes from the dipole limit $l=l'=1$, the
dependence on ${\bf \hat q}$ has the form
$({\bf\hat q}\cdot{\bf \hat R}_1)({\bf \hat q}\cdot{\bf \hat R}_N)$.
This form emphasizes paths beginning or ending in the direction ${\bf \hat q}$,
and acts like a ``search-light" in probing the structure of a system.
For polycrystalline materials or for measurements averaged over all ${\bf \hat q}$,
this dependence averages out.
Similarly, the contributions from the higher angular momenta couplings
probe other symmetries.

By defining a ${\bf q}$-dependent {\it effective scattering amplitude} 
$f_{\rm eff}({\bf q},k)$ one can use the path expansion (Eq.~(\ref{newform})) 
to express the  fine structure $\chi_{\bf q}(k)$ in a form identical to the
original EXAFS equation,
\begin{equation}
\chi_{\bf q}(k)= s_0^2\,\sum_{\Gamma} \,{\vert{f_{\rm eff}({\bf q}, k)}\vert\over kR^2}
\sin(2kR+\Phi_k) e^{-2 R/\lambda_k} e^{-2\sigma ^2 k^2},
\end{equation}
where $k=\sqrt{2(\omega+E_i)}$, $R=(1/2)\Sigma_i R_i$ is the
{\it effective path length}, and the prefactor $s_0^2\approx 0.9$ is a many-body
amplitude factor which accounts for inelastic losses (satellite excitations)
beyond the quasi-particle approximation\cite{campbell02}.
Here the effects of thermal and structural disorder are included using
a configurational average of Eq.~(\ref{newform}). These damping effects can be
approximated by a Debye-Waller factor $\exp(-2\sigma^2k^2)$, where $\sigma^2$
is the correlated mean-square variation $\langle (\delta R)^2\rangle$
for each scattering path \cite{rehr00}, while anharmonic terms
from the 1st and 3rd cumulants are generally weaker and contribute to the
phase.  Typically $\sigma^2$ is of order $10^{-2}$--$10^{-3}$ \AA$^2$.
For the case of full multiple-scattering, these effects can be calculated
to a good approximation by including a similar Debye-Waller factor
$\exp(-\sigma^2k^2)$ in
each propagator $G_{L{\bf R},L'{\bf R'}}(k)$.  Close to an absorption
edge where $k$ is small, the Debye-Waller factors are of order unity
and can be ignored.  In practical 
calculations $\sigma^2(T)$ can be approximated to reasonable accuracy using
the correlated Debye model evaluated for the Debye temperature
$\Theta_D$ of a material \cite{rehr00}.

\section{Examples}

\subsection{Be $K$-edge}

As mentioned above the momentum transfer magnitude $q$ and 
directional ${\bf\hat q}$ dependence of the $K$-edge in Be metal was 
measured and analyzed by Sternemann et al.~\cite{sternemann03}.
They found that the main changes in the spectra with increasing $q$ could be
explained by the increasing contribution from monopole ($s$-to-$s$)
type excitations.
This leads to a net decreasing anisotropy, i.e. decreasing dependence
on ${\bf\hat q}$ of the edge, since $s$-type states naturally have
no directional dependence. This conclusion is supported by ground state
local density of states calculations using full-potential
linear-augmented plane-wave and excited state calculations
using the BSE method of Ref.~\cite{soininen2001a}. 
\begin{figure}
\includegraphics[width=1.00\linewidth]{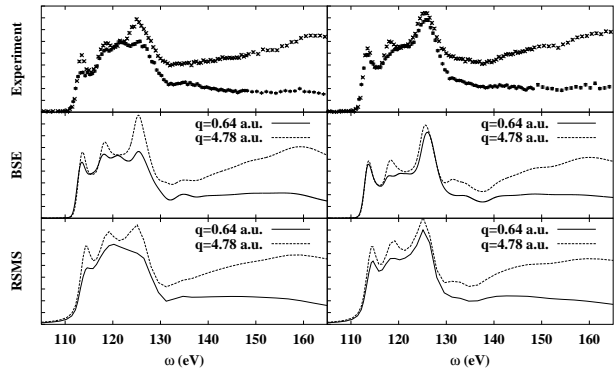}
\caption{Comparison between experiment~\protect\cite{sternemann03} (top
panels)
and inelastic loss spectra calculated using the BSE (middle panels)
and the RSMS method of this work (lower panels)
for the Be $K$-edge. The spectra are shown for ${\bf q}$ perpendicular to
the $c$-axis (left panels),
and ${\bf q}$ along the $c$-axis (right panels). Spectra are
shown for two values of momentum transfer $q$ as indicated in the
figure labels. The theoretical spectra was shifted by
4.5 eV to align with the experimental edge.}
\label{EvsF}
\end{figure}

In Fig.~\ref{EvsF}
we compare the inelastic scattering spectra calculated with the 
approach of this paper for the Be $K$-edge with the experimental results of
Ref.~\cite{sternemann03}.  From this 
comparison it is clear that our present calculation is capable of reproducing
both the momentum transfer direction and magnitude dependence of the
dynamic structure factor. However, there are noticeable discrepancies in
some fine details in the spectra within the first 15 eV. 
For the momentum transfer perpendicular to the $c$-axis, the 
120 eV features are  more pronounced in the RSMS calculation than in the 
experiment for both the low- and high momentum-transfer. However,
the main differences between these spectra is well reproduced by present
calculation. In the middle panel of Fig.~\ref{EvsF} we show for comparison the 
results from the BSE method~\cite{soininen2001a,sternemann03}. Considering 
how sensitive the problem of calculating core-excitation is, the agreement
between the two different theoretical methods is rather good.
This serves to validate the one-electron, final state rule calculations,
at least for this case. As noted above, most of the differences occur
rather close to the edge, which is also
the region most sensitive to the details of the core hole-electron
interaction and to the nature of the scattering potentials.  Beyond 
the first 15~eV the agreement between the two methods is very good. For 
${\bf q}$ along the $c$-axis, agreement between the two calculations is also
rather good for the energy range shown. For ${\bf q}$ perpendicular to 
the $c$-axis, however, the comparison is somewhat mixed. For the first peak
in the spectra it appears that RSMS accurately predicts the experimental
result.  The difference is that the first peak becomes a shoulder 
at low momentum transfer and a peak at higher momentum transfer.
On the other hand the details of the spectra
between 120-130~eV appear to be slightly better
reproduced by the BSE.   

\begin{figure}
\includegraphics[width=0.85\linewidth]{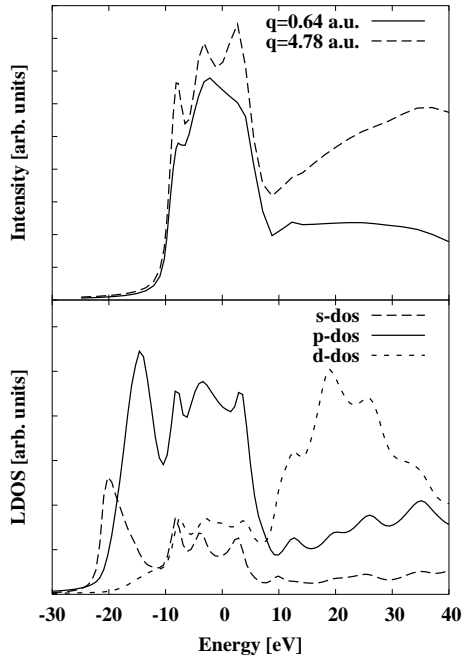}
\caption{Local projected DOS for Be together with
inelastic scattering spectra for two values of momentum transfer $q$
along the $x$-axis.  The Fermi level is at -9.4 eV.}
\label{BeLDOS}
\end{figure}

\begin{figure}
\includegraphics[width=1.00\linewidth]{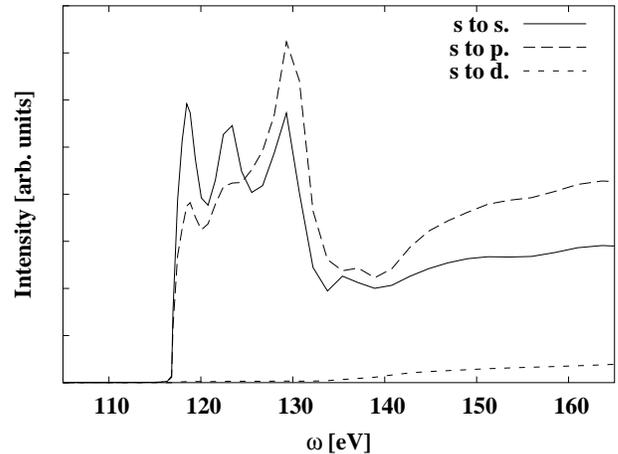}
\caption{Comparison of contributions to the inelastic loss spectra
from monopole and dipole transitions, for momentum transfer
$q=4.78$ a.u.\ along the $c$-axis.}
\label{SvP}
\end{figure}

As noted in Sec. II. C., the momentum transfer dependent changes in the 
spectra can be understood in terms of the changing weights of the different 
components of $\rho_{L,L'}(E)$ in the single crystal case, and the
LDOS $\rho_l(E)$ for polycrystalline systems.  
In Fig.~\ref{BeLDOS} we show the $s$- $p$- and $d$-DOS
for the core excited state at the site of the 
core hole in Be metal. Also shown are the calculated inelastic scattering
spectra for two values of momentum transfer along the $x$-axis. The calculated 
spectra are shifted so that the Fermi-energy is aligned with that
obtained in the LDOS calculation, where $E=0$ is the vacuum level.
On this scale the Fermi energy for Be as calculated by our RSMS code
is $E_F = - 9.4$ eV. The $p$-type LDOS close to the Fermi energy 
is composed of two peaks at $-8$ eV and $+4$ eV and a broad shoulder 
at $-4$ eV. In the small momentum transfer regime the dipole limit is 
reached, and the general shape of the $p$-type LDOS clearly shown in 
the shape of the calculated spectrum.The $s$-type LDOS on the other 
hand is composed of three sharp peaks at $-8$, $-4$ and $+3$ eV. Compared 
to the low $q$ spectrum, the high momentum transfer spectrum has sharp 
peaks at these same energies. The coincidence of the peaks is due to 
hybridization between the various angular
momentum components.  Clearly the changes in the near edge structure with
increasing $q$ can be attributed partly to the increasing contribution 
from the $s$-to-$s$ type transitions. Also the changes due to the matrix
elements that weight the different excitations have to be considered when 
comparing LDOS to experimental spectra. This can clearly  be
seen when comparing low momentum transfer spectra along either
the $x$- or $c$-axes. Although both exhibit a $p$-type LDOS structure,
the weights of the different features are strongly dependent on the
magnitude and direction 
of the momentum transfer. Also the overlap of the different types ($s$ vs $p$)
of LDOS is clearly demonstrated in these results.
The relative contributions 
to the sum of Eq.~(\ref{diagn}) 
from monopole ($s$-to-$s$), dipole 
($s$-to-$p$) transitions and quadrupole ($s$-to-$d$) are illustrated in
Fig.~\ref{SvP} for the higher momentum transfer $q=4.78$ a.u.\ along the 
$c$-axis. At lower momentum transfer the contributions from the monopole 
transitions and quadrupole transitions are negligible. At this 
relatively high momentum transfer the dipole forbidden $s$-to-$s$
transition is comparable in magnitude to the dipole allowed $s$-to-$p$
transition. 
The $s$-to-$d$ contribution is small at low energies but becomes increasingly 
important as the energy is increased, again reflecting the energy dependent 
weight of different excitation channels.  

\begin{figure}
\includegraphics[width=0.85\linewidth]{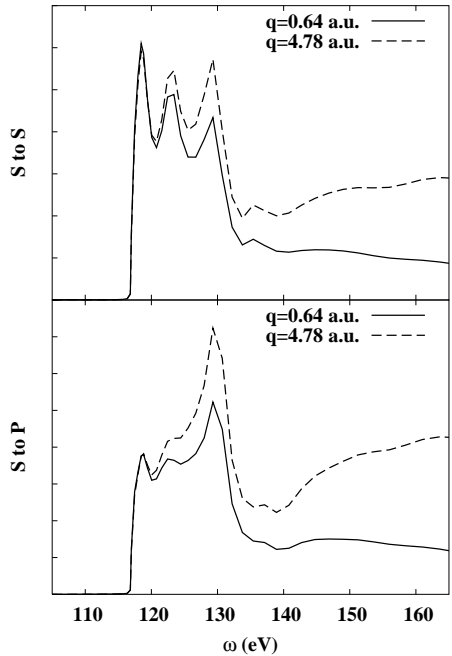}
\caption{Changes in the shape of different contributions to the
Be $K$-edge spectra, as a function of the magnitude of the 
momentum transfer along the $c$-axis for two values of $q$
listed in the figure labels.
The upper panel shows the changes for the monopole contribution
and the lower panel, those for the dipole allowed transitions.
The spectra were normalized so that the height of the
first peak is the same for both values of $q$.}
\label{qdecomp}
\end{figure}

The total spectrum changes also as function of the magnitude of the 
momentum transfer because of the changes in the channel specific 
spectra. This change is due to the excitation energy $E$ dependence 
of the final state partial wave $R_L({\bf r},E)$. As the momentum transfer 
changes the matrix elements
$M_L({\bf q}, E)=\langle R_L(E)| e^{i\bf q\cdot {\bf r}} |i\rangle$
make the channel (i.e., $L$) specific spectra change at an
energy $E$ dependent rate. 
In Fig.~\ref{qdecomp} the momentum transfer dependence of the dipole allowed
and monopole (dipole forbidden) transitions are examined separately.
The spectra are scaled so that the first peak for both momentum transfers 
is the same height. Besides this energy independent scaling factor there 
is also a momentum transfer dependence on the shape of the channel specific 
spectra. At a qualitative level this change is mostly visible in the 
central atom contribution. The momentum transfer dependent changes in the 
fine structure are less noticeable. 

\subsection{Al $L_1$ edge}
\begin{figure}
\includegraphics[width=0.95\linewidth]{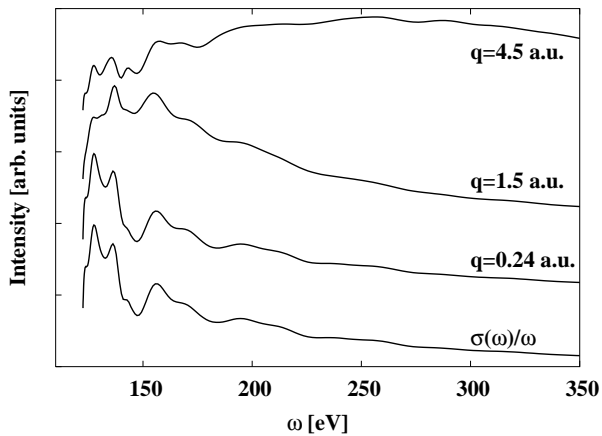}
\caption{Comparison of inelastic scattering and x-ray absorption spectra
at the Al $L_1$-edge using the path expansion for several
values of momentum transfer as indicated in the figure labels;
the curve $\sigma(\omega)/\omega$ is the result for $q=0$.
For clarity, the inelastic scattering spectra was slightly shifted
vertically. All the spectra where scaled to be shown on the same figure.
}
\label{AlExafs}
\end{figure}
As a final example we examine the core excited states in fcc Al. The calculated
x-ray absorption cross-section and the dynamic structure factor for $L_1$-edge at different
momentum transfers are compared in Fig.~\ref{AlExafs}.
To make contact with $S({\bf q},\omega)$ the absorption cross-section
$\sigma_{abs}(\omega)$ was divided by the excitation energy,
since in the limit ${\bf q}\rightarrow 0$,
\cite{mizuno67}
\begin{equation} 
\sigma_{abs}(\omega)\propto \omega S({\bf q},\omega).
\end{equation} 
 Comparison of 
$\sigma_{abs}(\omega)/\omega$ and $S({\bf q},\omega)$ for $q=0.24$ a.u.\
clearly shows the well known and experimentally verified~\cite{suzuki67}
fact that inelastic scattering results at low momentum transfer
can be used to obtain the same EXAFS information as in XAS.
Since we are interested in discussing the general 
trends of the momentum transfer dependence of $S_i({\bf q},\omega)$
in the extended energy range we have not included the Al $L_{2,3}$-edge
(edge located at about 80~eV) to this theoretical demonstration. For 
comparison with momentum transfer dependent experiments for this 
energy range (we are not aware of such experiments in any material) one 
would need to include these edges.    
As $q$ is increased to $q=1.5$ and $4.5$ a.u., the overall
shape of the spectra changes. As mentioned above, this change is mostly
due to changes in the central atom contribution. Although the 
EXAFS-like oscillations appear to diminish with higher momentum 
transfer, most of the broad oscillatory structure remains.   

\begin{figure}
\includegraphics[scale=0.7]{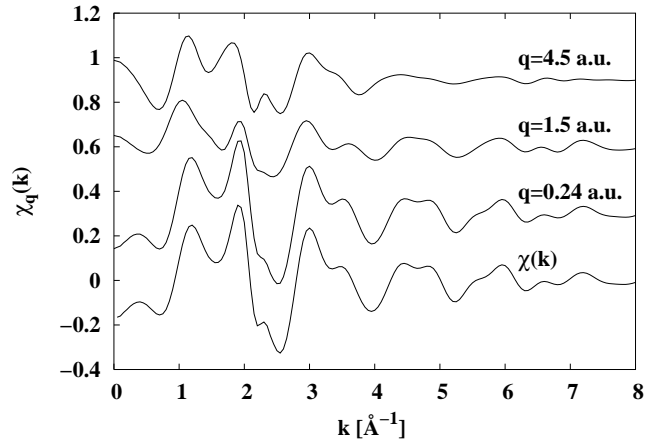}
\caption{Fine structure $\chi_{\bf q}(k)$ for Al at the $L_1$ edge
for different values of momentum transfer as listed in the figure labels,
together with the x-ray absorption fine structure $\chi(k)$, i.e., $q=0$.
The curves
are shifted for clarity.}
\label{Alchiq}
\end{figure}
\begin{figure}
\includegraphics[width=1.00\linewidth]{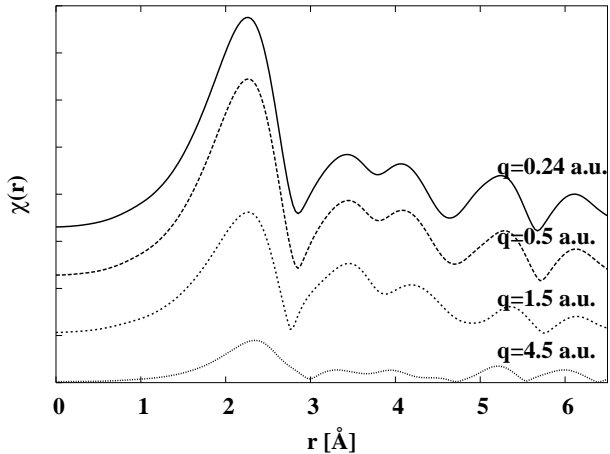}
\caption{Fourier Transform of Al fine structure for several values
of momentum transfer as listed in the figure labels.  The differences
reflect the interference between different channels contributing to
the spectra.
}
\label{AlExafsFT}
\end{figure}

In Fig.~\ref{Alchiq} we examine the momentum transfer dependence 
of the ${\bf q}$-dependent fine structure $\chi_{{\bf q}}(k)$.
Here we show the XAS fine structure $\chi(k)$,
together with $\chi_{{\bf q}}(k)$ for several values of $q$
at the Al $L_1$-edge. Again it is clear that for low $q$
the XAS result is reproduced. When the momentum transfer is increased, however,
the shape of the fine structure of the 
spectra is modified in a way that cannot be explained by simple scaling 
factors, due to the mixture between the various angular momentum contributions.
The changes are perhaps strongest close to the edge (at small $k$), but 
even the high $k$ spectra is modified. 
This suggests that it may be desirable to decompose 
the spectra into the various angular momentum components $\chi_l(k)$ prior 
to additional analysis. This is further illustrated in Fig.~\ref{AlExafsFT},
which shows the Fourier transform of the EXAFS
$\chi_q$ in $R$-space, phase corrected by the dominant central atom $p$-wave
phase shift $\exp(2i\delta_1)$.
Such Fourier transforms have peaks close to the near-neighbor distances
and provide a characterization of the near neighbor radial structure. 
Thus they can be used for quantitative fits of EXAFS in $R$-space.
Note that for small $q$ the transform is
insensitive to $q$ but becomes substantially more complex for the
larger values due to the overlapping contributions from the various
channels.

\section{Conclusions}

 We have shown how the real-space multiple-scattering (RSMS) approach
can be applied to model non-resonant inelastic scattering from deep core
levels for arbitrary condensed systems over a broad spectral range.
The approach is a generalization
to finite momentum transfer
of that used to model XAS based on the independent electron approximation
and the final state rule. In contrast to XAS where dipole selection
rules apply, couplings to all angular momentum components are important
and hence can probe different symmetries of the excited states.
Comparison of our approach with 
earlier work~\cite{soininen2001a,sternemann03} based on the particle-hole
BSE generally gives good agreement between these two
methods.  This suggests that an effective one-electron treatment that includes
the core-hole via the final state rule may be adequate for calculations
of inelastic scattering in addition to XAS.  We have also discussed how the
results may be used to analyze the inelastic x-ray (XRS) or electron 
scattering (EELS) from core electrons.  In particular we discussed the relation
between these spectra and the angular momentum projected density of
states (e.g., $s$-type and $p$-type LDOS) and density matrix components
$\rho_{L,L'}(E)$, and how these can be extracted from a series of experimental
measurements.  Our calculated spectra are
compared with the experimental XRS results~\cite{sternemann03} for the case 
of the Be $K$-edge, and gives good agreement over a wide energy
range, both for the directional and magnitude dependence of the 
momentum transfer. 
In addition we have shown that the calculations reproduce the well known
and demonstrated relationship between 
XAS and non-resonant inelastic scattering at low momentum transfers, i.e.,
the dipole limit $(q\rightarrow 0)$. An explicit example is given
for the case of the Al $L_1$-edge. 
Finally we have also discussed the momentum transfer dependence of
the fine structure $\chi_{\bf q}(k)$ in the EXAFS region, and suggested
how this might be exploited in structural studies.

\section{Acknowledgments}
We thank K. H\"am\"al\"ainen, G. Hug, 
E.~L. Shirley, 
S. Manninen, 
and J. Seidler for useful discussions. 
We would also like to thank C. Sternemann for discussions and for giving
us the experimental data for Be K edge in numerical form.
This work was supported in part by the U. S. Department of Energy grant
DE-FG06-97ER45623 and by NIST, and was facilitated by the DOE
Computational Materials Sciences Network. 
J.A.S was supported by
the  Academy of Finland (Grant No. 201291/40732).

\end{document}